\documentclass[12pt]{decar-wsd}

\usepackage{decar-common}
\usepackage{decar-post}

\title{Discretization of Linear Systems using the Matrix Exponential}
\author{Steven Dahdah and James Richard Forbes}

\begin{document}

\maketitle

\begin{abstract}
    \noindent Discretizing continuous-time linear systems typically requires numerical
    integration. This document presents a convenient method for discretizing the
    dynamics, input, and process noise state-space matrices of a continuous-time
    linear system using a single matrix exponential.
\end{abstract}

\section{Introduction}

Consider the continuous-time linear system,
\begin{align}
    \mbfdot{x}(t)
    &=
    \mbf{A}
    \mbf{x}(t)
    +
    \mbf{B}
    \mbf{u}(t)
    +
    \mbf{L}
    \mbf{w}(t),
    \\
    \mbf{y}(t)
    &=
    \mbf{C}
    \mbf{x}(t)
    +
    \mbf{M}
    \mbf{v}(t),
\end{align}
where $\mbf{x}(0) = \mbf{x}_0$ is the initial condition,
$\mbf{w}(t)$ is a random process representing the process noise, and
$\mbf{v}(t)$ is a random process representing the measurement noise. Both
processes are zero-mean Gaussian white noise, and are wide-sense stationary. As
such, their respective autocorrelation functions are
\begin{align}
    \ev{\mbf{w}(t)\mbf{w}(\tau)^\trans}
    =
    \mbf{Q}
    \delta(t - \tau),
    \\
    \ev{\mbf{v}(t)\mbf{v}(\tau)^\trans}
    =
    \mbf{R}
    \delta(t - \tau),
\end{align}
where $\delta(\cdot)$ is the Dirac delta function. The power spectral densities
of $\mbf{w}(t)$ and $\mbf{v}(t)$ are $\mbf{Q}$ and $\mbf{R}$ respectively.
Furthermore, the two random processes are uncorrelated. That is,
$\ev{\mbf{w}(t)\mbf{v}(\tau)^\trans} = \mbf{0}, \forall t \geq \tau$.

The corresponding discrete-time state-space system is
\begin{align}
    \mbf{x}_k
    &=
    \mbf{A}^{\!\mathrm{d}}
    \mbf{x}_{k-1}
    +
    \mbf{B}^\mathrm{d}
    \mbf{u}_{k-1}
    +
    \mbf{w}_{k-1}^\mathrm{d},
    &\mbf{w}_k^\mathrm{d} \sim \mathcal{N}(\mbf{0}, \mbf{Q}^\mathrm{d}),
    \\
    \mbf{y}_k
    &=
    \mbf{C}^\mathrm{d}
    \mbf{x}_k
    +
    \mbf{M}^\mathrm{d}
    \mbf{v}_k^\mathrm{d},
    &\mbf{v}_k^\mathrm{d} \sim \mathcal{N}(\mbf{0}, \mbf{R}^\mathrm{d}),
\end{align}
where $\mbf{x}_0 = \mbf{x}(0)$ is the initial condition,
$\mbf{A}^{\!\mathrm{d}}$, $\mbf{B}^\mathrm{d}$, $\mbf{C}^\mathrm{d}$, and
$\mbf{M}^\mathrm{d}$ are the discrete-time state-space matrices, and
$\mbf{Q}^\mathrm{d}$ and $\mbf{R}^\mathrm{d}$ are the covariance matrices of the
random variables $\mbf{w}_k^\mathrm{d}$ and $\mbf{v}_k^\mathrm{d}$. Note that
$\mbf{L}$ has been absorbed into $\mbf{Q}^\mathrm{d}$.
For more details about the discretization of linear systems with random inputs,
see~\cite[\S1.4.19,\S4.2.1,\S4.3.1]{barshalom_2001_estimation}.

The matrices $\mbf{C}^\mathrm{d}$, $\mbf{M}^\mathrm{d}$, and
$\mbf{R}^\mathrm{d}$ are straightforward to compute.
Since the measurement equation has no dynamics,
\begin{align}
    \mbf{C}^\mathrm{d} &= \mbf{C},
    \\
    \mbf{M}^\mathrm{d} &= \mbf{M}.
\end{align}
The measurement noise covariance matrix is~\cite[\S3.2.1]{groves}
\begin{gather}
    \mbf{R}^\mathrm{d}
    =
    \frac{1}{\Delta t}
    \mbf{R},
\end{gather}
where $\Delta t$ is the sampling period.
The matrices $\mbf{A}^{\!\mathrm{d}}$, $\mbf{B}^\mathrm{d}$, and
$\mbf{Q}^\mathrm{d}$ are typically computed using
\begin{align}
    \mbf{A}^{\!\mathrm{d}}
    &=
    \exp(\mbf{A} \Delta t),
    \\
    \mbf{B}^\mathrm{d}
    &=
    \int_{0}^{\Delta t}
        \exp(\mbf{A} (\Delta t - s)) \,
        \mbf{B}
    \,\dee s,
    \\
    \mbf{Q}^\mathrm{d}
    &=
    \int_{0}^{\Delta t}
        \exp(\mbf{A}(\Delta t - s)) \,
        \mbf{L} \mbf{Q} \mbf{L}^\trans
        {\exp(\mbf{A} (\Delta t - s))}^\trans
    \,\dee s,
\end{align}
where $\exp(\cdot)$ is the matrix exponential.

Note that, in practice, $\mbf{R}^\mathrm{d}$ is often specified directly instead
of discretizing $\mbf{R}$. This is not possible for $\mbf{Q}^\mathrm{d}$, since
the process noise passes through a differential equation. Its discretized form
is therefore dependent on the differential equation as well as the power
spectral density $\mbf{Q}$.

In this document, an alternative method to compute $\mbf{A}^{\!\mathrm{d}}$,
$\mbf{B}^\mathrm{d}$, and $\mbf{Q}^\mathrm{d}$ using a single matrix exponential
is presented.
This method is based on Farrell~\cite[\S4.7.2.1]{farrell}, where a method to
compute $\mbf{A}^{\!\mathrm{d}}$ and $\mbf{Q}^\mathrm{d}$ using the matrix
exponential is described. Cited in Farrell~\cite[\S4.7.2.1]{farrell}
is a method from
Van~Loan~\cite{vanloan}, that also computes $\mbf{B}^\mathrm{d}$ in the same
step.
However, Van~Loan does not use the same definition of the $\mbf{Q}^\mathrm{d}$
matrix as Farrell. To resolve this inconsistency, a modified version of
Van~Loan's method is presented here. Using this method, it is possible to
compute the $\mbf{A}^{\!\mathrm{d}}$, $\mbf{B}^\mathrm{d}$, and
$\mbf{Q}^\mathrm{d}$ using a single matrix exponential.
A version of this approach also appears in Barfoot~\cite[\S C.3]{barfoot}.

\section{Discretization method}

To compute $\mbf{A}^{\!\mathrm{d}}$, $\mbf{B}^\mathrm{d}$, and
$\mbf{Q}^\mathrm{d}$ with sampling period $\Delta t$, first construct the matrix
\begin{align}
    \mbs{\Xi} &=
    \bma{cccc}
        \mbf{A} & \mbf{L} \mbf{Q} \mbf{L}^\trans & \mbf{0} & \mbf{0} \\
        \mbf{0} & -\mbf{A}^\trans & \mbf{0} & \mbf{0} \\
        \mbf{0} & \mbf{0} & \mbf{A} & \mbf{B} \\
        \mbf{0} & \mbf{0} & \mbf{0} & \mbf{0}
    \ema.
\end{align}
Multiplying by the sampling period and taking the matrix exponential yields
\begin{align}
    \mbs{\Upsilon}
    &= \exp(\mbs{\Xi} \, \Delta t) \nonumber \\
    &= \bma{cccc}
        \mbs{\Upsilon}_{11} & \mbs{\Upsilon}_{12} & \star   & \star \\
        \mbf{0}             & \star               & \star   & \star \\
        \mbf{0}             & \mbf{0}             & \star   & \mbs{\Upsilon}_{34} \\
        \mbf{0}             & \mbf{0}             & \mbf{0} & \star
    \ema,
\end{align}
where $\star$ represents the (potentially) nonzero sub-matrices that are
unnecessary.

The discrete-time state-space matrices can be calculated as
\begin{align}
    \mbf{A}^{\!\mathrm{d}} &= \mbs{\Upsilon}_{11}, \\
    \mbf{B}^\mathrm{d} &= \mbs{\Upsilon}_{34}, \\
    \mbf{Q}^\mathrm{d} &= \mbs{\Upsilon}_{12} \mbs{\Upsilon}_{11}^\trans.
\end{align}

\section{Proof}

The foundation of Van~Loan's method is \Cref{th1}, which is taken directly
from~\cite{vanloan}.

\begin{theorem}
    Let $n_1$, $n_2$, $n_3$, and $n_4$ be positive integers, and set $m$ to be
    their sum. If the $m \times m$ block triangular matrix $\mbs{\Xi}$ is
    defined by
    \begin{align}
        \mbs{\Xi}
        &=
        \bma{cccc}
            \mbf{A}_1 & \mbf{B}_1 & \mbf{C}_1 & \mbf{D}_1 \\
              \mbf{0} & \mbf{A}_2 & \mbf{B}_2 & \mbf{C}_2 \\
              \mbf{0} & \mbf{0}   & \mbf{A}_3 & \mbf{B}_3 \\
              \mbf{0} & \mbf{0}   & \mbf{0}   & \mbf{A}_4 \\
        \ema
        \begin{array}{c}
            \} \, n_1 \\
            \} \, n_2 \\
            \} \, n_3 \\
            \} \, n_4 \\
        \end{array} \\[-3ex]
        &\phantom{{}={}}
        \phantom{\Bigg[}
        \begin{array}{cccc}
            \phantom{\mbf{A}_1} &
            \phantom{\mbf{B}_1} &
            \phantom{\mbf{C}_1} &
            \phantom{\mbf{D}_1} \\
            \rotatebox[origin=c]{90}{\{} &
            \rotatebox[origin=c]{90}{\{} &
            \rotatebox[origin=c]{90}{\{} &
            \rotatebox[origin=c]{90}{\{} \\[-1ex]
            n_1 & n_2 & n_3 & n_4 \\
        \end{array}
        \phantom{\Bigg]}\nonumber
    \end{align}
    then for $t \geq 0$
    \begin{align}
        \exp(\mbs{\Xi}t)
        &=
        \underbrace{
        \bma{cccc}
            \mbf{F}_1(t) & \mbf{G}_1(t) & \mbf{H}_1(t) & \mbf{K}_1(t) \\
                 \mbf{0} & \mbf{F}_2(t) & \mbf{G}_2(t) & \mbf{H}_2(t) \\
                 \mbf{0} & \mbf{0}      & \mbf{F}_3(t) & \mbf{G}_3(t) \\
                 \mbf{0} & \mbf{0}      & \mbf{0}      & \mbf{F}_4(t) \\
        \ema
    }_{\mbs{\Upsilon}(t)}
    \end{align}
    where
    \begin{align}
        \mbf{F}_j(t)
        &=
        \exp(\mbf{A}_j t), \quad j \in \{1,2,3,4\}, \\
        \mbf{G}_j(t)
        &=
        \int_0^t
            \exp(\mbf{A}_j(t-s)) \, \mbf{B}_j \exp(\mbf{A}_{j+1} s)
            \,\dee s, \quad j \in \{1,2,3\}, \\
        \mbf{H}_j(t)
        &=
        \int_0^t
            \exp(\mbf{A}_j (t-s)) \, \mbf{C}_j \exp(\mbf{A}_{j+2} s)
        \,\dee s \nonumber \\
        &\quad + \int_0^t \int_0^s
            \exp(\mbf{A}_j (t-s)) \, \mbf{B}_j
            \exp(\mbf{A}_{j+1} (s-r)) \, \mbf{B}_{j+1}
            \exp(\mbf{A}_{j+2}r)
            \,\dee r \, \dee s, \quad j \in \{1,2\},
    \end{align}
    and
    \begin{align}
        \mbf{K}_1(t)
        &=
        \int_0^t
            \exp(\mbf{A}_1(t-s)) \, \mbf{D}_1 \exp(\mbf{A}_4s)
        \,\dee s \nonumber \\
        &\quad + \int_0^t \int_0^s
            \exp(\mbf{A}_1(t-s))
            \left(
                \mbf{C}_1\exp(\mbf{A}_3(s-r)) \, \mbf{B}_3
                + \mbf{B}_1\exp(\mbf{A}_2(s-r)) \, \mbf{C}_2
            \right)
            \exp(\mbf{A}_4 r)
        \,\dee r \,\dee s \nonumber \\
        &\quad + \int_0^t \int_0^s \int_0^r
            \exp(\mbf{A}_1(t-s)) \, \mbf{B}_1
            \exp(\mbf{A}_2(s-r)) \, \mbf{B}_2
            \exp(\mbf{A_3}(r-w)) \, \mbf{B}_3
            \exp(\mbf{A}_4w)
        \,\dee w \,\dee r \,\dee s.
    \end{align}\label{th1}
\end{theorem}

By choosing the entries of $\mbs{\Xi}$ carefully, the matrices
$\mbf{A}^{\!\mathrm{d}}$, $\mbf{B}^\mathrm{d}$, and $\mbf{Q}^\mathrm{d}$ can be
computed easily from the elements of $\mbs{\Upsilon}(\Delta t)$ using
\Cref{th1}.

First, consider the definition of $\mbf{Q}^\mathrm{d}$,
\begin{align}
    \mbf{Q}^\mathrm{d}
    &=
    \int_{0}^{\Delta t}
        \exp(\mbf{A}(\Delta t - s)) \,
        \mbf{L} \mbf{Q} \mbf{L}^\trans
        {\exp(\mbf{A} (\Delta t - s))}^\trans
    \,\dee s. \label{eq:qd}
\end{align}

Rearranging \cref{eq:qd} as
\begin{align}
    \mbf{Q}^\mathrm{d}
    &=
    \int_{0}^{\Delta t}
        \exp(\mbf{A}(\Delta t - s)) \,
        \mbf{L} \mbf{Q} \mbf{L}^\trans
        \exp(-\mbf{A}^\trans s)
    \,\dee s \;
    {\exp(\mbf{A} \Delta t)}^\trans
\end{align}
yields a similar form to
\begin{align}
    \mbf{G}_1(\Delta t)
    &=
    \int_0^{\Delta t}
        \exp(\mbf{A}_1(\Delta t-s)) \, \mbf{B}_1 \exp(\mbf{A}_2 s)
    \,\dee s.
\end{align}

If $\mbf{A}_1 = \mbf{A}$, $\mbf{B}_1 = \mbf{L}\mbf{Q}\mbf{L}^\trans$, and
$\mbf{A}_2 = -\mbf{A}^\trans$, then
\begin{align}
    \mbf{Q}^\mathrm{d} &= \mbf{G}_1(\Delta t) \; {\exp(\mbf{A} \Delta t)}^\trans.%
    \label{eq:qd_result}
\end{align}

Next, consider the definition of $\mbf{A}^{\!\mathrm{d}}$,
\begin{align}
    \mbf{A}^{\!\mathrm{d}}
    &=
    \exp(\mbf{A} \Delta t). \label{eq:ad}
\end{align}

\Cref{eq:ad} has the same form as
\begin{align}
    \mbf{F}_1(\Delta t)
    &=
    \exp(\mbf{A}_1 \Delta t).
\end{align}

Thus with the same choice of $\mbf{A}_1$,
\begin{align}
    \mbf{A}^{\!\mathrm{d}} &= \mbf{F}_1(\Delta t).
\end{align}

Furthermore, \cref{eq:qd_result} simplifies to
\begin{align}
    \mbf{Q}^\mathrm{d} &= \mbf{G}_1(\Delta t) \, {\mbf{F}_1(\Delta t)}^\trans.
\end{align}

Finally, consider the definition of $\mbf{B}^\mathrm{d}$,
\begin{align}
    \mbf{B}^\mathrm{d}
    &=
    \int_{0}^{\Delta t}
        \exp(\mbf{A} (\Delta t - s)) \,
        \mbf{B}
    \,\dee s. \label{eq:bd}
\end{align}

\Cref{eq:bd} has a similar form to
\begin{align}
    \mbf{G}_3(\Delta t)
    &=
    \int_0^{\Delta t}
        \exp(\mbf{A}_3(\Delta t-s)) \, \mbf{B}_3 \exp(\mbf{A}_4 s)
    \,\dee s.
\end{align}

If $\mbf{A}_3=\mbf{A}$, $\mbf{B}_3=\mbf{B}$, and $\mbf{A}_4=\mbf{0}$, then
\begin{align}
    \mbf{B}^\mathrm{d}
    &=
    \mbf{G}_3(\Delta t).
\end{align}

Note that choosing $\mbf{G}_2(\Delta t)$ instead of $\mbf{G}_3(\Delta t)$ is
not possible, as it would lead to conflicting choices of $\mbf{A}_2$.

Thus, by applying \Cref{th1}, the choices
\begin{align}
    \mbf{A}_1 &= \mbf{A}, \\
    \mbf{B}_1 &= \mbf{L}\mbf{Q}\mbf{L}^\trans, \\
    \mbf{A}_2 &= -\mbf{A}^\trans, \\
    \mbf{A}_3 &= \mbf{A}, \\
    \mbf{B}_3 &= \mbf{B}, \\
    \mbf{A}_4 &= \mbf{0},
\end{align}
allow $\mbf{A}^{\!\mathrm{d}}$, $\mbf{B}^\mathrm{d}$, and $\mbf{Q}^\mathrm{d}$
to be computed as
\begin{align}
    \mbf{A}^{\!\mathrm{d}} &= \mbf{F}_1(\Delta t), \\
    \mbf{B}^\mathrm{d} &= \mbf{G}_3(\Delta t), \\
    \mbf{Q}^\mathrm{d} &= \mbf{G}_1(\Delta t) \, {\mbf{F}_1(\Delta t)}^\trans,
\end{align}
using a single matrix exponential operation.
\qed

\printbibliography

\end{document}